\documentclass[conference]{IEEEtran}
\usepackage[utf8]{inputenc}
\usepackage{graphicx}
\usepackage{caption}
\usepackage{subfig}

\usepackage{epsfig}
\usepackage[T1]{fontenc}
\usepackage{textcomp}
\usepackage{amsmath,amssymb}
\usepackage{cite}
\usepackage{array}
\usepackage{tabularx}
\usepackage{color}
\usepackage{algorithm,algorithmic}

\begin{document}

\title{On the topology effects in wireless sensor networks based prognostics and health management}
\author{\IEEEauthorblockN{Ahmad Farhat, Abdallah Makhoul, and Christophe Guyeux}
\IEEEauthorblockA{FEMTO-ST Laboratory, DISC department\\
University of Franche-Comt\'e\\
Rue Engel-Gros, 90016 Belfort, France\\
}
\and
\IEEEauthorblockN{Rami Tawil, Ali Jaber, and Abbas Hijazi}
\IEEEauthorblockA{Department of Computer Science\\
Lebanese University\\
Beirut, Lebanon\\
}
}

\maketitle

\begin{abstract}
In this work, we consider the usage of wireless sensor networks (WSN) to monitor an area of interest, in order to diagnose on real time its state. Each sensor node forwards information about relevant features towards the sink where the data is processed. Nevertheless, energy conservation is a key issue in the design of such networks and once a sensor exhausts its resources, it will be dropped from the network. This will lead to broken links and data loss. It is therefore important to keep the network running for as long as possible by preserving the energy held by the nodes. Indeed, saving the quality of service (QoS) of a wireless sensor network for a long period is very important in order to ensure accurate data. Then, the area diagnosing will be more accurate. From another side, packet transmission is the phase that consumes the highest amount of energy comparing to other activities in the network. Therefore, we can see that the network topology has an important impact on energy efficiency, and thus on data and diagnosis accuracies. In this paper, we study and compare four network topologies: distributed, hierarchical, centralized, and decentralized topology and show their impact on the resulting estimation of diagnostics. We have used six diagnostic algorithms, to evaluate both prognostic and health management with the variation of type of topology in WSN.
\end{abstract}

\section{Introduction}

Due to the increasing demand in reliability and quality of service, modern industrial plants witness a continuously growing complexity. As a result, the costs of failure and system downtime are getting more expensive. Therefore, monitoring these areas is very essential to evaluate their health and diagnose them at any time, and then to plan maintenance activities to avoid disastrous failure results. Prognostic and Health Management (PHM) is a process that allows an advanced system to automatically test the area, diagnose it, isolate the failure, and try to predict the Remaining Useful Life (RUL) for an area before failure takes place~\cite{sun2007research}. By doing so, a maintenance scheduling is then determined and the area shutdown is prevented. It is worth mentioning that if the prediction model and the provided measurements are not accurate, there is a high possibility that the maintenance activity will not be done on time.

Health assessment and diagnostics activity of the area, that is followed by prediction of RUL, requires online measurements of the operating conditions of the area of interest. These information are usually gathered through a number of sensor nodes. In this study, we consider the case where sensors communicate their information within a Wireless Sensor Network (WSN). WSNs are different from traditional computer networks, as the former are composed by a large number of sensor nodes with very limited and non-renewable energy. Most of the time, they are deployed to capture the occurrence of possible events in hostile and inaccessible areas~\cite{yick2008wireless}. A classical assumption in PHM is that monitoring data is available and complete, which is not always true. Due to the nature of communication in this network and to the characteristics of its devices, a WSN is at risk of failure so this will have an effect on the accuracy and completeness of the data that will be captured, and consequently on PHM. Therefore, one of our objectives is to maintain the Quality of Service (QoS) of WSN as long as possible to ensure the accuracy of the data of the monitored area. If such issue is not taken into consideration while building a PHM process over a WSN, the provided results of diagnostic or prognostic may not be reliable.

From several factors and important parameters in WSN like: lifetime, security, data aggregation, packet transfer, density, etc, the topology in WSN has an important impact on the accuracy of data and then on PHM. The variability of network topologies due to
node failures, introduction of additional nodes, variations in sensor location, requires the adaptability of underlying network structures and operations. From another side, in order to save more energy sensor nodes may be activated or deactivated into scheduling mechanisms in order of keeping as much as possible a dense coverage and achieve fault tolerance.
%
%
%
%
Thus, the diagnostic processes must be compatible with these strategies, and with a device's coverage of a changing quality.

In this paper, we study the topologies in WSN and its relation with prognostic and health management. We focus on the impact of topologies on the accuracy of the data captured by the wireless sensor network, and its consequences on the diagnostic of the status of the monitored area. 
Our objective is to show that usual diagnostic processes that perform well in classical data provided by a well deployed wired network of sensors, may face a dramatic decrease of performances in the case where data are obtained via a WSN, due to the diversity and variation of topologies. To do so, we used six machine learning algorithms to diagnose the area state, namely the so-called Support Vector Machines (SVM), Naive Bayes (NB), Random Forests (RF), Gradient Tree Boosting (GTB), Tree-Based Feature Selection (TBFS), and Nearest Neighbors (NN) methods. In addition, we study four different types of topology (the most used in WSN) which are: distributed, hierarchical, centralized, and decentralized topology. 
 
The remainder of this article is structured as follows. Section~\ref{sec:Topologies in wireless sensor networks} presents an overview of WSNs topologies. In Section~\ref{sec:Prognostic and health management and its linkage with WSN topologies}, we detail the links that can be established between PHM and the topologies in WSN field or research. We simulate and describe four different topologies in WSN to show their impact on diagnostics, and the results of these simulations are given in Section~\ref{sec:Numerical simulations}. This article ends with a conclusion section, where the contribution is summarized and intended future work is provided.

\section{Topologies in wireless sensor networks}
\label{sec:Topologies in wireless sensor networks}

In wireless sensor networks, the connectivity of the network is established via radio
transmission between sensors. For two sensors to be able to communicate, they must be within some critical range of each other, as transmission capability is finite. A network is connected if any node can communicate with any other node, possibly using intermediate nodes as relays. The variability of network topologies (connectivity) due to node failures, introduction of additional nodes, variations in sensor location, requires the
adaptability of underlying network structures and operations. Since sensors may be spread in an arbitrary manner, One of the fundamental issues that arises in sensor networks in addition to the connectivity is the coverage. In order to ensure connectivity and data accuracy in addition to coverage, WSN use redundant coverage where multiple sensors nodes cover the same physical location. Therefore, coverage may vary across the network. A solution to save energy in the network rises
on finding scheduling mechanisms. The objective of such mechanisms is to activate or deactivate redundant nodes while keeping as much as possible a dense coverage and ensuring connectivity.

Another metric to save energy in sensor networks is to reduce the amount of data collected and transmitted via the network. Data gathering in WSNs can be either periodic or event-driven~\cite{mak4}. In periodic applications~\cite{mak2, mak3}, data is gathered periodically while in event-driven applications gathering depends on the occurrence of some events. In both cases, the goal from aggregation is reducing energy dissipation by holding packets for as long as possible in intermediate nodes. All packets will be combined together then forwarded in the network. It is obvious to see that a decrease in energy consumption leads to an increase in the overall delay, and vice versa. A reliable solution would aim at finding an acceptable trade off between energy consumption and delay in WSNs \cite{mak1, Kwon11}.

WSNs can be either heterogeneous or homogeneous~\cite{li2008survey}. In the latter, all nodes have the same role and characteristics. In the former, nodes have different roles: some nodes simply sense and forward information while others aggregate data, manage their area, perform computations, etc. Consequently, some of the nodes can be equipped
with higher energy, longer radio range, etc. Several WSN topologies were used in existing monitoring applications, but all of them revolved around four different types (or models) of topologies which are: distributed, hierarchical, centralized, or decentralized topology.  

\begin{itemize}

\item \textbf{Distributed topology:} in distributed topologies, there is no management of the network by the central node (or a region of it). They consist of a collection of nodes having equal roles. Therefore, no aspect of hierarchy is considered. No prior infrastructure is imposed before the network starts running; each node discovers its surrounding area and decides which node (or nodes) to communicate with. This decision usually relies on the radio range and the transfer distance. Distributed topologies render the network’s maintenance an easy task: if a node fails, its neighbors, within their sensing range, will establish new links with other nodes, and the network will continue to work normally.

\item \textbf{Hierarchical topology:} the organization of sensor nodes can be in several levels, making a hierarchical topology (or a tree topology). Level $0$ is represented by the root and there is no level above. From two adjacent levels, sensor nodes are connected in an end to end manner. The hierarchical model can be seen as three different layers: $(1)$ the core layer (the root), which is enhanced for availability and performance, $(2)$ the distribution layer, which implements policies and forwards messages, and $(3)$ the access layer (the leaf nodes), which represents the access point to the network. Scalability is the advantage of Hierarchical WSN. The network is more manageable and the task of isolating and detecting faults is simplified due to the presence of different levels.

\item \textbf{Centralized topology:} its one of the easiest topologies to design and implement (also called star topology). All the sensor nodes have a simple task which is sensing new information and forwarding it to a central node where all the data processing will be proceeded with. One of the major problems of this topology is that it presents a single point of failure. The whole network will become paralyzed if a problem occurs at the central node: The data packet cannot be forwarded nor processed when a new event is detected.

\item \textbf{Decentralized topology: }decentralized topologies are considered as a combination of the distributed and the centralized topologies. The network is divided into regions (or clusters) which are locally managed by a central node (called the Cluster Head CH). This topology offers a reasonable settlement between energy consumption and Quality of Service (QoS). In this type of topology, there is a reduction of congestion problem and the network no longer has a single point of failure.

\end{itemize}

\section{WSN topologies impact on prognostics and health management}

\label{sec:Prognostic and health management and its linkage with WSN topologies}

Maintenance is an important activity in industry which is either performed to revive a machine/component, or to prevent it from breaking down. It aims at increasing system availability, readiness and enhancing safety. Different strategies have evolved through time in order to bring maintenance to its current state: condition-based and predictive maintenance. The increasing demand of reliability in industry caused this evolution. PHM is a tool to predict the Remaining Useful Life (RUL) of engineering assets and is the key process of condition-based and predictive maintenance. Nowadays, industrial machines are required to avoid shutdowns while offering safety and reliability~\cite{peng2010current}. Research in PHM field has gained and was given a great deal of attention. Prognostic models are developed in an attempt to predict the RUL of machinery (or monitored area) before failure takes place. If there is no accuracy in the prediction model and the provided measurements, the maintenance activity will possibly be performed either too soon or too late.

\begin{figure}
\centering
\includegraphics[width=8cm,height=40mm]{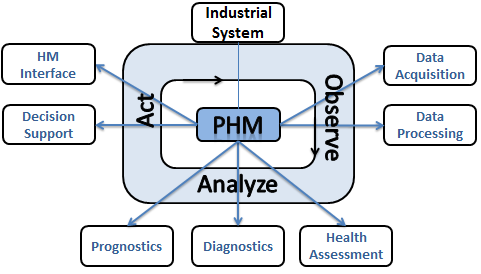}
\caption{The process of PHM.}
\label{fig:phm}
\end{figure}

Condition Based Maintenance (CBM) was proposed and developed in early nineties~\cite{heng2009rotating}, and it is based on real-time observations. It is an on-line approach that assesses machine’s health through condition measurements. As any maintenance strategy, CBM aims to increase the system reliability and availability while reducing costs of maintenance. This particular strategy has benefits which include avoiding unnecessary maintenance tasks and costs, as well as not interrupting the normal machine operations~\cite{heng2009rotating}. CBM decreases the number of maintenance operations
and reduces the influence of human error. 
A new maintenance has recently emerged which is the Predictive maintenance (PM). It predicts the system health in the future and defines the needed maintenance activities accordingly based on the current condition.
Shifting from traditional maintenance strategies to CBM and PM requires extra tasks. These tasks encompass data analysis and modeling, system surveillance, and decision making support system. This scientific approach is called PHM.
PHM is the core activity of CBM and PM. The steps of PHM are: data acquisition, data processing, health assessment, diagnostics, prognostics, and decision making support~\cite{jardine2006review}, this is done following the steps described in Figure~\ref{fig:phm}. 

The aim of diagnostics is to specify and quantify an actual failure while the aim of 
prognostics is anticipating failures. Prognostics estimate the RUL by considering the past events, in addition to the machine’s current state, and operating conditions~\cite{jardine2006review}. By studying the evolution of continuous measurements of parameters that need to be tracked in time to assess the machine’s state, this estimation is done. These parameters can be temperature, humidity, vibration, pressure, and so on. There is a fixed threshold for the monitored parameter. Once this threshold is reached, an alarm goes on indicating that a symptom of system deteriorating has been detected. After that, a diagnosis of the state of the system is made and the RUL is computed with an associated confidence limit. There are two causes for the uncertainties of the RUL predictions: either the threshold value of monitored parameter, or the RUL prediction itself. The necessary prerequisites for reliable prognostics are proposed in~\cite{monitoring1diagnostics}.

Reliability is necessary in industry (monitored area in general). 
For the past years, the research in prognostics resulted in variety of tools and techniques that offer the possibility for plants to survey their systems, anticipate failures, and schedule maintenance activities. WSNs are mainly designed for surveillance purposes. They can be deployed in many fields such as military, automotive, agriculture, medicine, and so on~\cite{li2008survey}. Recently, a great deal of attention was given to WSN applications by industry. These sensor networks are used to monitor their machinery for maintenance scheduling. Furthermore, data will be provided by the sensors deployed to survey the system/component in order to assess the health, diagnose the system, and estimate the RUL. However, inaccuracy in the data will cause the prediction based on it to be irrelevant. The topologies in WSN have important impact on the accuracy of data and therefore have an important impact on PHM. Before the network starts running, studying and choosing the topologies in WSNs need to be considered. The aim of this study is to reveal the impact of topologies in WSN on the accuracy of the captured data from the monitored area and therefore on PHM. We can say that the accuracy of the data is related to the topologies used in WSN from several factors and important parameters in WSN. Lifetime is one of the most important factors in WSN which is related to topologies, and this is because the costs related to energy consumption varies with the variation of topologies (data aggregation, packet transfer distance, frequency, etc). Security is another important factor also related to topologies, and this is because the role and characteristics of nodes differ according to topologies (some nodes simply sense and forward information while others aggregate data, manage their area, perform computations, etc). Several factors other than those mentioned play an important role in the accuracy of data in WSN with the variation of applications (topologies), such as density, batteries of nodes, data aggregation, etc. What is worth mentioning is that data aggregation is important in increasing the lifetime of network as mentioned before, but on the other hand, data aggregation always reduces the data accuracy, so the error rate of diagnosis is greatly related to the method of data aggregation. Since good predictions rely on real data, it is certain that the first step to be done in the research is ensuring a reliable source of information.

\section{Numerical study}
\label{sec:Numerical simulations}

\subsection{Experimental protocol}
\label{sec:Experimental protocol}

\subsubsection{WSN simulation}
\label{sec:Wireless sensor network simulation}

In this paper, in order to show the impact of WSN topologies on the PHM, we used three types of sensing fields: temperature, pressure, and humidity. Therefore, we considered a network of ~$300$ sensor nodes, sensing respectively the levels of temperature ($100$~sensors), pressure ($100$), and humidity ($100$~sensors).
%
%
Each sensor node has a battery of $300u$ ($u$ is the battery unit), and captures specific data depending on the operating age $t$. We consider that no level of correlation is introduced between the different features:

\begin{itemize}
\item Under normal conditions, temperature sensors follow a Gaussian law of parameter $(20\times (1+0.005t), 1)$, in case of a malfunction of the area in the range of this sensor, these parameters are mapped to $(350, 20)$. Finally, these sensors return the value~$2$ when they break down.

\item The pressure sensors produce data following a Gaussian law of parameter $(5\times (1 + 0.01t), 0.3)$ when they are sensing a well-functioning area. The parameters changed to $(20, 2.5)$ in case of area failure in the location where the sensor is placed, as long as the pressure sensors return~$1$ when they are broken down.

\item The Gaussian parameters are $(52.5\times (1 + 0.001t), 12.5)$ when both the area and the humidity sensors are in normal conditions. These parameters are set to $(80, 10)$ in case of area failure in the range of this sensor, whereas malfunctioning humidity sensors produce the value~$3$.
\end{itemize}

Each sensor follows a Poisson process ($Pp$) of parameter $(200\times (1-0.01t)+0.01)$, to determine if a breakdown occurs in the location where sensor is placed. Subsequently all of these sensors execute Algorithm~\ref{algo11}.

\begin{algorithm}
\caption{Sensor algorithm}
\label{algo11}
\begin{algorithmic}
\IF{$Pp < 1$}
    \STATE{the area and the sensors are in normal conditions}
       
\ELSE   
    \IF{$1 \le Pp < 100$}
        \STATE{the area in failure (in the range of this sensor)}
    \ELSE 
        \STATE{the sensor is broken down}
                 
    \ENDIF
       
\ENDIF 
\end{algorithmic}
\end{algorithm}

Each category of sensors has its own constant threshold, depending on the abnormality of the sensed data. If the captured data by the sensor in a specific category exceeded the threshold, this indicates that a symptom of system deteriorating has been detected. Then a diagnostic study aims for specifying and quantifying an actual failure (whether it failed or not). In this work, we used six algorithms for diagnosis, which are mentioned in Section~\ref{sec:Machine learning algorithms}. 
In this study, 
we consider the values of the thresholds as follows: $26$~degrees for temperature, $7$~bars in pressure, and $80$~percents of humidity.

The deployment strategy (manually or randomly) of sensors~\cite{patilissues}, the adjustment of the coverage range of sensors~\cite{zhu2012survey}, and the density in WSN~\cite{adlakha2003critical} have an important impact on the accuracy of the data captured by WSN that will be used to diagnose the state of area. In order to study the impact of topologies of WSN on diagnostics, in our simulation we consider the following:

\begin{itemize}

\item Most of the times, the area to be monitored is hazardous and hard to access because of the difficulty in its geographical area like monitoring the forests, oceans, military zones, etc. Therefore in this study, we used random deployment for area monitoring. 

\item Suppose that in this work, the region to be monitored is a rectangle of area $A = L \times W$ such that $L$ and $W$ are the length and width of the monitored region respectively. The area of the coverage range of the sensor is mostly related to the area of the region to be monitored. Therefore we consider the area of the coverage range to be equal~$1\%$ of the total area of the region. Subsequently, the coverage radius will be $R= 1/10 \times \sqrt{A/\pi}$.

\item Suppose that the density of sensors in the monitored area is constant ($300$~sensors), and that the area is fully covered by these sensors at time~$t=0$ (when the WSN starts working).

\end{itemize}

\subsubsection{Machine learning algorithms}
\label{sec:Machine learning algorithms}



The research in PHM is very broad and the authors working in this domain use several algorithms in order to perform the diagnostic of the state of system. These algorithms in literature are called machine learning algorithms. In machine learning, classification refers to identifying the class to which a new observation belongs, on the basis of a training set and quantifiable observations, known as properties. Machine learning displays a detailed study about the system and from it, an algorithm is built. These algorithms can be operated by building a model from examples inputs in order for the algorithm to be able to diagnose or take decision for new data. 

We have chosen six machine learning algorithms (to diagnose the system) which were used before by several authors in literature in order to evaluate the PHM. Our study in this work focused on evaluating these six diagnostic algorithms with the variation of the topology of WSN. These algorithms are: Support Vector Machine (SVM)~\cite{galar2012remaining}, Naive Bayes (NB)~\cite{zhang2004optimality, ng2014naive}, Random Forests (RF)~\cite{elghazel2015random, Breiman01}, Gradient Tree Boosting (GTB)~\cite{buhlmann2007boosting}, Tree-Based Feature Selection (TBFS)~\cite{sugumaran2007feature}, and Nearest Neighbors (NN)~\cite{mcroberts2012estimating}.

Finally we need a large and reliable data set in order to train these algorithms. So that, later, we can diagnose the system (area monitoring) from the new data that will be captured by WSN. For that, we take data consisting of $N$ lines, each line is composed by $T$ temperature data, $P$ data of pressure, and $H$ data of humidity to train these algorithms. All of these data are generated in the same way mentioned in Section~\ref{sec:Wireless sensor network simulation} (same type of data that will be captured by WSN during area monitoring).

\subsection{Simulation results}

In order to illustrate the impact of topologies on the quality of data and the diagnosing of the state of the monitored area, we simulated four different topologies: decentralized topology, distributed topology, hierarchical topology and centralized topology. 

\subsubsection{Decentralized topology}
\label{sec:Decentralized topology}

In order to study the impact of decentralized topology on the diagnosing of state of area, we consider that the nodes are grouped into $30$~clusters. Each cluster is managed by a leader called cluster head (CH) or aggregator which is equipped with batteries of $1500u$. The sensors capture the data from the area and send it to the CH, the latter aggregates the data and send it to another CH or to the sink. In this study, we consider that the data aggregation at the CH happens as follows: 

\begin{equation}
\sum_{i=0}^{S-1}D_i^c / S
\end{equation}

where $D$ is the data sent from the sensor to the aggregator, $c$ is the type of sensor (temperature, pressure, or humidity), and $S$ is the number of data that will be aggregated each time (for example, every~$3$ data from a certain type which are sent to CH from sensors, undergo aggregation).

\begin{figure}[ht]
 \centering
 \subfloat[Sensors network at time $t=0$.\label{decentraliser}]{
  \includegraphics[width=4cm,height=30mm]{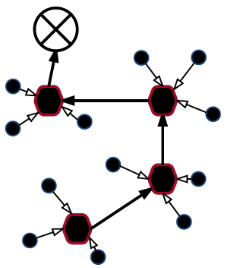}
  } 
  \hfill
 \subfloat[Sensors network at time $t=x$.\label{decentraliserX}]{
  \includegraphics[width=4cm,height=30mm]{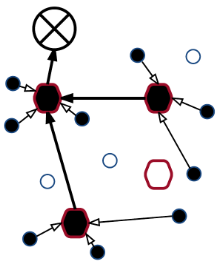}
  }
 \caption{Scenario of decentralized topology. 
 \label{Scenario of decentralized topology}} 
\end{figure}

The scenario of this topology is shown in Figure~\ref{Scenario of decentralized topology}, the deployment of sensors is random, and the distribution and partition of CH on sensors follows $K$-means clustering method. Each sensor sends data to its CH. The latter, after aggregating these data, sends it to the closest CH on the condition that this CH is closest to the sink. If no CH meets this requirement, it will send it directly to the sink as shown in Figure~\ref{decentraliser}. After time $t=x$, the CH and sensors may become inactive for several reasons most importantly energy consumption or activity scheduling. If a CH became inactive, sensors in this cluster find other closest clusters to be in. In addition, CHs communicating with this inactive CH change their routes to the closest active CH. 
%
%
%
%
What is worth mentioning is that the black circles are the active sensors, the white circles are the inactive sensors, the black hexagons are the active CH, the white hexagons are the inactive CH, and finally the crossed circle is the sink.

\begin{figure}[!ht]
\centering
\includegraphics[width=9cm,height=5cm]{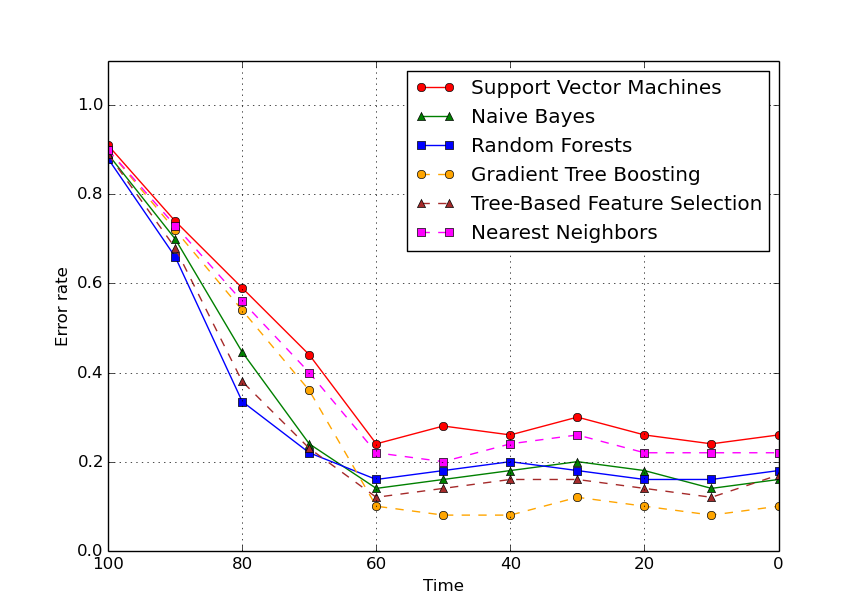}
\caption{Error rate in diagnostics if the topology is decentralized with the variation of the time.}
\label{err1}
\end{figure}

In this study, we supposed that the area is fully covered by these sensors after they have been randomly deployed. As mentioned before, the topology may be dynamic, the sensors or CHs on the long term will die (because of energy consumption) or break down (due to various causes as the operating age).
Figure~\ref{err1} shows the variation of error rate for the six considered algorithms, in the case where the topology is decentralized, with the variation of time $t$ (operating age). Each point in this figure is an average of error rates of a given algorithm on $20$~simulations (for a certain $t$). As shown in the figure, during $t=0$ | $t=60$ (if $0 \leq t \leq 60$), each algorithm has a specific error interval (in \%) as follows: $[24,30]$ for SVM, $[14,20]$ for NB, $[16,20]$ for RF, $[8,12]$ for GTB, $[12,17]$ for TBFS, and $[22,26]$ for NN. After that (if $t > 60$) the error rate for each algorithm increased significantly at these intervals to reach at $t=70$, $44~\%$ for SVM, $24~\%$ for NB, $22~\%$ for RF, $36~\%$ for GTB, $23~\%$ for TBFS, and $40~\%$ for NN. This shows that at this time the sensors and CH in WSN are dying or breaking down, and this fact leads to the presence of uncovered places in the area (coverage hole) and therefore incomplete data for diagnostics. Then when the WSN exceeds $t=60$ (if $t>60$) the error rate of algorithms increase as time increase to reach $91~\%$ at $t=100$ if the algorithm is SVM, $89~\%$ if NB, $88~\%$ if RF, $90~\%$ if GTB, $89~\%$ if TBFS, and $90~\%$ if NN (approximately the whole network is inactive). What is worth mentioning is that the error rate in this simulation (decentralized topology) is related to the method of data aggregation.

\subsubsection{Distributed topology}
\label{sec:Distributed topology}

\begin{figure}[ht]
\centering
 \subfloat[Sensors network at time $t=0$.\label{distibuer}]{
  \includegraphics[width=4cm,height=30mm]{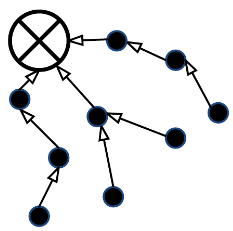}
  }
 \hfill
 \subfloat[Sensors network at time $t=x$.\label{distibuerX}]{
  \includegraphics[width=4cm,height=30mm]{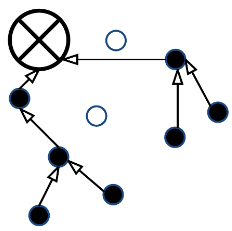}
  }
\caption{Scenario of distributed topology. 
\label{Scenario of distributed topology}}  
\end{figure}

%
The scenario of distributed topology is shown in Figure~\ref{Scenario of distributed topology}, where all sensor nodes in the network have the same role and importance; \textit{i.e.} there is no aggregation role, no clusters, and no CHs. Data packets are forwarded in a hop-by-hop manner. Each sensor is able to discover its neighbors within a radio range of $2R_c$ ($R_c$ is the coverage range). We assume that every node can access information about its neighbors, including their locations. The nodes choose neighbors to communicate with, and the latter should be closest to the sink within the sender's radio range.
If the sensor is closest to the sink, the sensor will then send it directly to the sink.

As explained in the scenario before, after certain time $t=x$, the sensors may become inactive and the routes always change in function of the closest neighbors to the sink as shown in Figure~\ref{distibuerX}. 
What is worth mentioning is that the black, white, and crossed circles represent the active sensors, inactive sensors, and the sink respectively.

\begin{figure}[!ht]
\centering
\includegraphics[width=9cm,height=5cm]{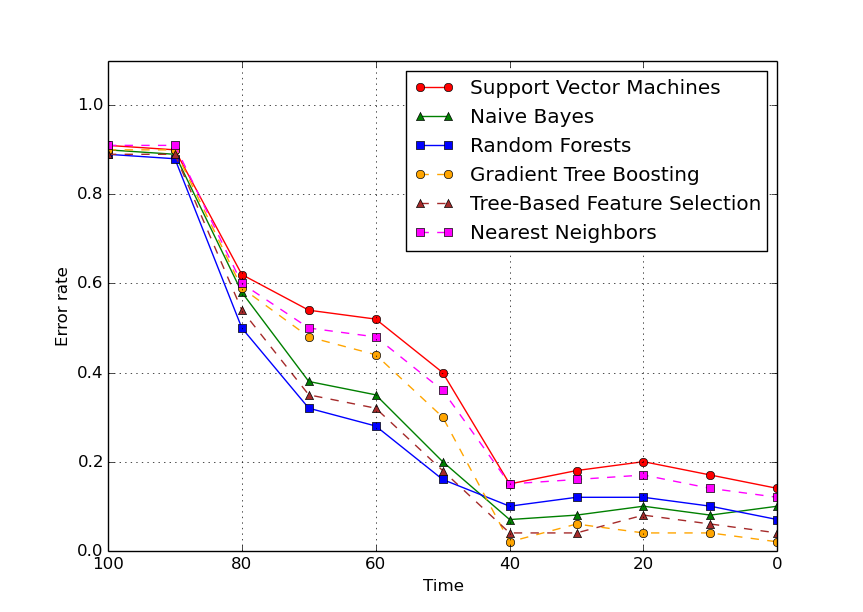}
\caption{Error rate in diagnostics if the topology is distributed with the variation of the time.}
\label{err2}
\end{figure}

Figure~\ref{err2} presents the variation of error rate for the six considered algorithms 
in the case of distributed topology, with the variation of time $t$ (operating age). Each point in this figure is an average of error rates of a given algorithm on $20$~simulations (for a certain $t$). As shown in the figure, during $t=0$ | $t=40$ (if $0 \leq t \leq 40$), each algorithm has a specific error interval (in \%) as follows: $[14,20]$ for SVM, $[7,10]$ for NB, $[7,12]$ for RF, $[2,6]$ for GTB, $[4,8]$ for TBFS, and $[12,17]$ for NN. After that (if $t > 40$) the error rate for each algorithm increased significantly at these intervals to reach at $t=50$, $40~\%$ for SVM, $20~\%$ for NB, $16~\%$ for RF, $30~\%$ for GTB, $18~\%$ for TBFS, and $36~\%$ for NN. This shows that at this time the sensors in WSN are dying or breaking down, and this fact leads to the presence of uncovered places in the area (coverage hole) and therefore incomplete data for diagnostics. Then when the WSN exceeds $t=40$ (if $t>40$) the error rate of algorithms increase as time increase to reach $90~\%$ at $t=90$ if the algorithm is SVM, $89~\%$ if NB, $88~\%$ if RF, $90~\%$ if GTB, $89~\%$ if TBFS, and $91~\%$ if NN (approximately the whole network is inactive).

\subsubsection{Hierarchical topology}
\label{sec:Hierarchical topology}

As we have mentioned in Section~\ref{sec:Topologies in wireless sensor networks}, sensor nodes can be organized in several levels, making a hierarchical topology. The sensor nodes are organized in a tree hierarchy from the sink (being the root of a tree), until sensor nodes having no descendants (leaf nodes). 
In order to study the impact of hierarchical topology on diagnosis, and to compare this topology to other topologies, we took WSN composed of $300$ sensors (leaf nodes), and each sensor has $300u$ for battery. 
These sensors are considered as the access layer in this topology (third layer).
We considered $30$ nodes playing the role of the second layer in topology (the distribution layer), which implements policies and forward messages. These nodes responsible for building the links between the leaf nodes towards the sink (core layer). Each sensor from these $30$ has $300u$ for battery supply, on the contrary, in a decentralized topology, CHs are given an extra supply, therefore the batteries last longer and we dispose with more data for diagnostics.

\begin{figure}[ht]
\centering
 \subfloat[Sensors network at time $t=0$.\label{hierarchie}]{
  \includegraphics[width=4cm,height=30mm]{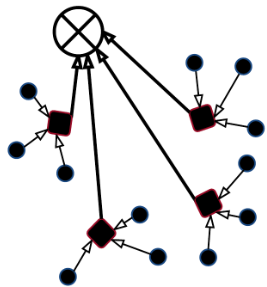}
  }
 \hfill
 \subfloat[Sensors network at time $t=x$.\label{hierarchieX}]{
  \includegraphics[width=4cm,height=30mm]{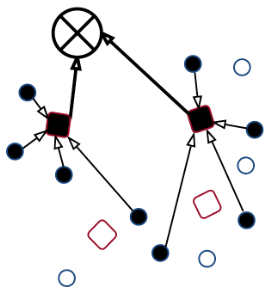}
  }
\caption{Scenario of hierarchical topology. 
\label{Scenario of hierarchical topology}}  
\end{figure}

The scenario of this topology is shown in Figure~\ref{Scenario of hierarchical topology}.
After a certain time $t=x$, the sensors in access layer or distribution layer may become inactive for several reasons most importantly energy consumption. Unfortunately if a parent node (in distribution layer) become inactive, its children can no longer communicate with other nodes in the network. In this case, in order to keep connectivity, each sensor will then communicate with the closest active node in the distribution layer as shown in Figure~\ref{hierarchieX}.

\begin{figure}[!ht]
\centering
\includegraphics[width=9cm,height=5cm]{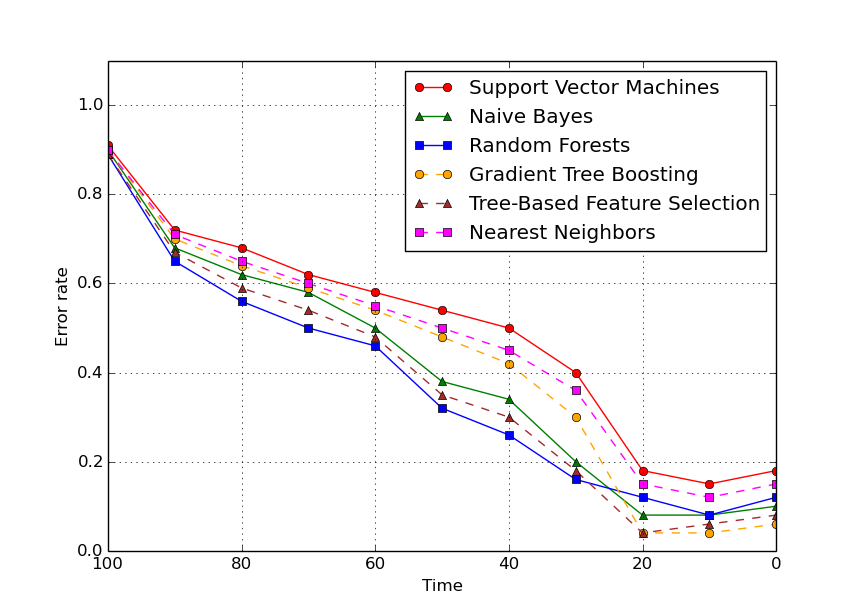}
\caption{Error rate in diagnostics if the topology is hierarchical with the variation of the time.}
\label{err3}
\end{figure}

Figure~\ref{err3} indicates the variation of error rate for the six considered algorithms under the same conditions of the previous studies, but here it is the case where the topology is hierarchical with the variation of time $t$. Each point in this figure, is an average of error rates of a given algorithm on $20$~simulations (for a certain $t$). As shown in the figure, during $t=0$ | $t=20$ (if $0 \leq t \leq 20$), each algorithm has a specific error interval (in \%) as follows: $[15,18]$ for SVM, $[8,10]$ for NB, $[8,12]$ for RF, $[4,6]$ for GTB, $[4,8]$ for TBFS, and $[12,15]$ for NN. These intervals are approximately the same where the topology is distributed (where the whole network is active), and this is because in these two topologies, there is no data aggregation as the decentralized topology. After that (if $t > 20$) the error rate for each algorithm increased significantly at these intervals to reach at $t=30$, $40~\%$ for SVM, $20~\%$ for NB, $16~\%$ for RF, $30~\%$ for GTB, $18~\%$ for TBFS, and $36~\%$ for NN. This shows that at this time the sensors in WSN (in access or distribution layer) are dying or breaking down, and this fact leads to the presence of uncovered places in the area (coverage hole) and therefore incomplete data for diagnostics. Then when the WSN exceeds $t=20$ (if $t>20$) the error rate of algorithms increase as time increase to reach $91~\%$ at $t=100$ if the algorithm is SVM, $90~\%$ if NB, $89~\%$ if RF, $90~\%$ if GTB, $89~\%$ if TBFS, and $90~\%$ if NN (approximately the whole network is inactive).

\subsubsection{Centralized topology}
\label{sec:Centralized topology}

\begin{figure}[ht]
\centering
 \subfloat[Sensors network at time $t=0$.\label{centraliser}]{
  \includegraphics[width=4cm,height=30mm]{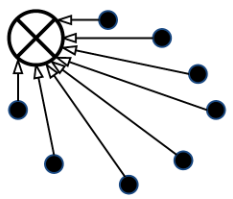}
 } \hfill
 \subfloat[Sensors network at time $t=x$.\label{centraliserX}]{
  \includegraphics[width=4cm,height=30mm]{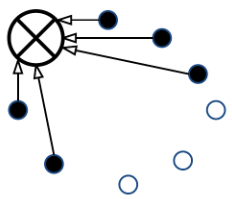}
 }
\caption{Scenario of centralized topology. 
\label{Scenario of centralized topology}}  
\end{figure}

In centralized topology, all the sensor nodes have the simple task of sensing new information and forwarding it to a central node where all the data processing is done as shown in Figure~\ref{centraliser}. 
In this topology, we can notice that after $t=x$, the nodes that exhaust their energy first are the farthest from the sink. This is due to the long distance of packet transfer as shown in Figure~\ref{centraliserX}. The black and white circles are the active and inactive sensors respectively, and the crossed circle is the sink.

\begin{figure}[!ht]
\centering
\includegraphics[width=9cm,height=5cm]{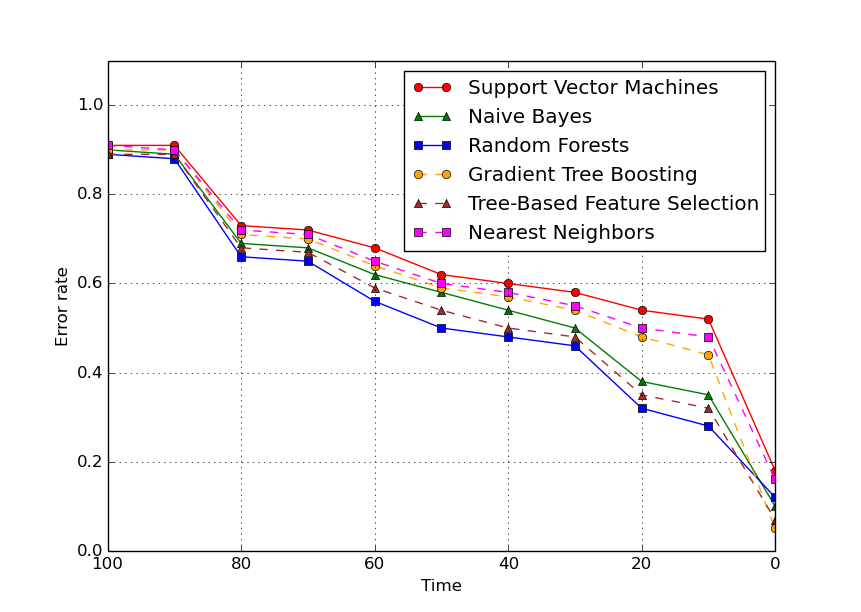}
\caption{Error rate in diagnostics if the topology is centralized with the variation of the time.}
\label{err4}
\end{figure}

Figure~\ref{err4} shows the variation of error rate for the six considered algorithms under the same conditions of the previous studies. Each point in this figure is an average of error rates of a given algorithm on $20$ simulations (for a certain t). As shown in the figure, at $t=0$ (when the WSN starts working) each algorithm has a specific error rate (in \%) as follows: $18~\%$ for SVM, $10~\%$ for NB, $12~\%$ for RF, $5~\%$ for GTB, $7~\%$ for TBFS, and $16~\%$ for NN. During the work of the network, with time, the sensors that are located farthest from the sink start dying first because they consume more energy than the others, and this is due to the long distance of packet transfer. For that at $t=10$ the error rate increased in a noticeable way to become $52~\%$ with SVM, $35~\%$ with NB, $28~\%$ with RF, $44~\%$ with GTB, $32~\%$ with TBFS, and $48~\%$ with NN. This is a proof that the data became incomplete for diagnostic, and this is  because the regions far from the sink are no longer covered by sensors. Then when the WSN exceeds $t=10$ (if $t>10$) the error rate of algorithms increase as time increase to reach $91~\%$ at $t=90$ if the algorithm is SVM, $89~\%$ if NB, $88~\%$ if RF, $90~\%$ if GTB, $89~\%$ if TBFS, and $90~\%$ if NN (approximately the whole network is inactive).

\subsubsection{Discussion}
\label{sec:discussion}

In this section we will explain and compare the results we obtained in our study in order to focus on several issues or parameters related to topologies which have a great impact on diagnostics (PHM). We notice in Figure~\ref{err2} (with distributed topology), the noticeable variation of error rate of the algorithms with time from what is shown in Figure~\ref{err1} (with decentralized topology).
We note from these two figures that in Figure~\ref{err2} the sensors after $t=40$ ($t>40$) started dying or breaking down, and the whole network became inactive at $t=90$, while in Figure~\ref{err1} the sensors or CH started dying or breaking down after $t=60$ ($t>60$), and the whole network became inactive at $t=100$. Moreover we can notice that in Figure~\ref{err1}, during $t=0$ | $t=60$ (the whole network is active) the error rate is evolving in a way larger than the one in Figure~\ref{err2} during $t=0$ | $t=40$ (where the whole network is active). From this study and based on this comparison we can conclude that the lifetime of the networks with decentralized topology is greater than if it were distributed topology, and this is due to that the data aggregation reduces the number of packet transfer, and therefore it further reduces the overall energy consumption in the network. But the error rate of diagnosis is greatly related to the method of data aggregation (if the topology is decentralized) because data aggregation always reduces the data accuracy that will be used to diagnose, and this is shown and clarified in these two figures where the whole network is active.

We had a different scene in Figure~\ref{err3} (with hierarchical topology) because the variation of error rate of the algorithms varied with time in a significant way from what is shown in Figure~\ref{err1} and~\ref{err2} (with decentralized and distributed topology). We notice from these figures that in Figure~\ref{err3} the sensors after $t=20$ ($t>20$) started dying or breaking down, while in previous studies, the sensors became inactive after this time as shown in the figures (variation of error rate with time). Based on this study, we can conclude that the lifetime of the network with hierarchical topology is smaller than if it were decentralized or distributed topology (the network lifetime defined as time until the first node dies). Furthermore, we note that the whole network with hierarchical topology became inactive at $t=100$ as the network with decentralized topology, while with distributed topology the whole network became inactive at $t=90$. Based on these results, we can conclude the importance of dividing the WSN in area monitoring into regions which are locally managed by a central node (or parent node).

If we suppose that the network lifetime can alternatively be defined as the time until the first node dies, then by relying on the change of curves in these figures, we conclude that the lifetime of the network with centralized topology is smaller than if it were decentralized, distributed, or hierarchical topology. Moreover, the whole network with hierarchical and decentralized topology became inactive at $t=100$, while with distributed and centralized topology the whole network became inactive at $t=90$. Based on these four results, we confirm what we mentioned before about the importance of dividing the WSN in area monitoring into regions which are locally managed by a parent node (the network remains active for a longer time, therefore the sink continues to receive information from the monitoring area for a longer time). Based on this work, we were able to notice the importance and impact of each type of topology in WSN on diagnostics with the increase of operating age of WSN, and focus on several issues related to these types of topologies.

\section{Conclusion and Future Work}

The WSNs provide a new way of distributed data collection and wireless transmission for PHM to diagnose the state of an area and to be informed if it is in failure or not. Topologies in WSN are important factors to achieve QoS in WSNs application. In this paper, we explained the relation between WSN topologies and their impact on PHM and the area diagnostic. We mentioned and studied four different topologies in WSN, each one of them belonging to a certain type as follows: distributed, hierarchical, centralized, and decentralized topology. 
In this work, we focused on several issues related to these types of topologies. We studied the lifetime of each type, and concluded that the lifetime of decentralized topology is larger than the other. Therefore we can say that this type of topology is the best for PHM reliability, and this is because the complete data from the area are available for a longer time. 
We conclude also that dividing the WSN in area monitoring into regions which are locally managed by a parent node (like decentralized and hierarchical topology) is very important, because the network in this case remains active for a longer time. As a future work, we plan to study other factors like density, data aggregation, coverage and scheduling, communication, etc and their impact on PHM.


\textit{This work is  partially funded by the Labex ACTION program (contract ANR-11-LABX-01-01)}

\bibliographystyle{plain}

\bibliography{references}
\end{document}